\documentclass[12pt]{article}
\usepackage[english]{babel}
\usepackage{graphicx,epsfig}
\makeindex \textwidth 170mm \textheight 220mm \topmargin -5mm
\newcommand{\be}{\begin{equation}}
\newcommand{\ee}{\end{equation}}
\newcommand{\bea}{\begin{eqnarray}}
\newcommand{\eea}{\end{eqnarray}}

\def\celmec{{Celest. Mech.}}
\def\far{{Farinella}}
\def\mtc#1{\mathcal{#1}}
\def\jgr{{J. Geophys. Res.}}
\def\yse{Yarkovsky-Schach effect}
\def\re{Rubincam effect}
\def\asas{{Astron. Astrophys.}}
\def\med#1{<{#1}>_{2\p}}
\def\sv{\hat s}
\def\sa{semimajor axis}
\def\ogf{(\omega +f)}
\def\vok{{Vokrouhlick{\'{y}}}}

\def\mashn{Mashhoon}

\def\rfr#1{eq.(\ref{#1})}
\def\rfrs#1#2{eqs.(\ref{#1})-(\ref{#2})}
\def\Rfr#1{Eq.(\ref{#1})}

\def\bb{\bibitem}
\def\eqi{\begin{equation}}
\def\eqf{\end{equation}}
\def\eqia{\begin{eqnarray}}
\def\eqfa{\end{eqnarray}}

\def\rp#1#2{{#1\over#2}}

\def\lb#1{\label{#1}}
\def\lg{LAGEOS}

\def\grm{gravitomagnetism}

\def\p{\pi}

\def\f{\phi}

\begin{document}
\begin{titlepage}
\begin{flushright}
\today\\
BARI-TH/00\\
\end{flushright}
\vspace{.5cm}
\begin{center}
{\LARGE Satellite non-gravitational orbital perturbations and the
detection of the  gravitomagnetic clock effect } \vspace{1.0cm}
\quad\\
{Lorenzo Iorio$^{\dag}$\\
\vspace{0.5cm}
\quad\\
{\dag}Dipartimento di Fisca dell' Universit{\`{a}} di Bari, via
Amendola 173, 70126, Bari, Italy\\} \vspace{1.0cm}
 {\bf Abstract\\}
\end{center}

{\noindent \small The general relativistic gravitomagnetic clock
effect consists in the fact that two massive test bodies orbiting
a central spinning mass in its equatorial plane along two
identical circular trajectories, but in opposite directions, take
different times in describing a full revolution with respect to an
asymptotically inertial observer. In the field of the Earth such
time shift amounts to 10$^{-7}$ s. Detecting it by means of a
space based mission with artificial satellites is a very demanding
task because there are severe constraints on the precision with
which the radial and azimuthal positions of a satellite must be
known: $\delta r\sim 10^{-2}$ cm and $\delta \varphi\sim 10^{-2}$
milliarcseconds per revolution. In this paper we assess if the
systematic errors induced by various non-gravitational
perturbations allow to meet such stringent requirements. A couple
of identical, passive laser-ranged satellites of \lg\ type with
their spins aligned with  the Earth's one is considered. It turns
out that all the non vanishing non-gravitational perturbations
induce systematic errors in $r$ and $\varphi$ within the required
constraints for a reasonable assumption of the mismodeling in some
satellite's and Earth's parameters and/or by using dense
satellites with small area-to-mass ratio. However, the error in
the Earth's $GM$ is by far the largest source of uncertainty in
the azimuthal location which is affected at a level of 1.2
milliarcseconds per revolution.}

{\noindent \small }
\end{titlepage} \newpage \pagestyle{myheadings} \setcounter{page}{1}
\vspace{0.2cm} \baselineskip 14pt

\setcounter{footnote}{0}
\setlength{\baselineskip}{1.5\baselineskip}
\renewcommand{\theequation}{\mbox{$\arabic{equation}$}}
\noindent
\section{Introduction}
General Relativity, in its weak-field and slow-motion
approximation, predicts that two massive test particles following
in opposite directions two geodesic circular orbits of identical
radius in the equatorial plane of a central spinning body take
different times in describing a full revolution with respect to an
asymptotically inertial observer. If $t^{+}$ denotes the period of
the particle moving in the same sense of the rotation of the
central body and $t^{-}$  denotes the period of the particle
moving in the opposite sense, it turns out that  \eqi\Delta
t=t^{+}-t^{-}=4\pi \frac{J}{Mc^{2}},\lb{mash}\eqf where $J$ and
$M$ are the proper angular momentum and the mass, respectively, of
the central source and $c$ is the speed of light in vacuum. Note
that the Newtonian gravitational constant $G$ does not appear in
\rfr{mash}. The co-rotating particle turns out to be slower than
the counter-rotating one. This is the so called gravitomagnetic
clock effect [{\it Cohen and \mashn,} 1993; {\it Mashhoon et al.,}
1999; 2000].  \Rfr{mash} is also equal to the difference in the
test particles' proper periods $\tau^{\pm}$ for test particles
orbiting at distances larger than the gravitational radius
$r_{g}=2GM/c^{2}$ of the central body and neglecting terms of
order $\mathcal{O}(c^{-4})$.

For a couple of artificial Earth satellites \rfr{mash} yields a
time shift of $10^{-7}$ s which is large enough to be detected
according to the present-day level of accuracy in timing
measurements with atomic clocks [{\it Lichtenegger et al.,} 2000].
Note that \rfr{mash} is independent of the satellite's orbit
radius. This feature could be fruitfully exploited in choosing
suitably the orbital parameters of the satellites to be employed.
However, it has been  shown in [\textit{Gronwald et al.,} 1997;
\textit{Lichtenegger et al.,} 2000] that, in order to make
feasible such measurement, the radial and azimuthal positions of
the satellites should be known at a level of $\delta r\sim
10^{-2}$ cm and $\delta\varphi\sim 10^{-2}$ milliarcseconds (mas)
per revolution. Such constraints are very stringent: suffice it to
say that the position of \lg\ laser-ranged satellite is presently
known at a centimeter level. Recently, in [{\it Iorio,} 2001; {\it
Lichtenegger et al.,} 2001] the impact of the long-periodic and
short-periodic gravitational perturbations, with the systematic
errors induced by them in $r$ and $\varphi$, have been worked out:
the conclusion is that, at the present level of knowledge of the
static and time-varying parts of the Earth's gravitational field
[{\it Lemoine et al.}, 1998], the gravitational errors are larger
than the required $\delta r$ and $\delta \varphi$.

However, in the near future the new, more accurate data for the
terrestrial gravitational field from the CHAMP and GRACE missions
will be available and the situation could become more favorable.
So, it appears important to assess the error budget due to the
non-gravitational forces [{\it Milani et al.,} 1987] on the
satellite's radial and azimuthal locations, which is the scope of
this paper.

We will consider throughout it a couple of identical passive,
geodetic, laser-ranged spherical satellites of \lg\ type both
because Satellite Laser Ranging (SLR)  has reached in the last
decade an astonishingly level of accuracy and because with this
kind of satellites  it is by far simpler to model accurately
enough the non-gravitational perturbations. Indeed, they depend on
the physical and geometrical characteristics of the satellites and
on the geometry of their orbits as well. For example, twice a
year, almost six months apart, the Sun, moving along the ecliptic,
intersects  the Earth's equatorial plane so that the satellite's
revolutions occurring these times are affected by the phenomenon
of eclipses. The radiative perturbations which are generated,
directly or indirectly, by the solar electromagnetic radiation
like the direct solar radiation pressure, the albedo and the \yse\
act differently on the satellite's orbit with respect to the
orbital arcs described in full sunlight and in many cases such
discrepancies are relevant and difficult to calculate. Moreover,
the thermal forces which cause the Yarkovsky-Schach and the
Rubincam effects depend on both the physical properties of the
satellite and on the orientation of its spin axis with respect to
an inertial frame. The terrestrial environment may cause the spin
direction to change in a more or less predictable way over time
spans of some years, as it is occurring to \lg\
[\textit{M\'{e}tris et al.,} 1999].

In order to detect the gravitomagnetic time shift, which is
cumulative, there is no need of very long orbital arcs: indeed,
for example, the orbital period of \lg\ amounts to $1\times
10^{-1}$ days only, so that $10^{2}-10^{3}$ revolutions
[\textit{Lichtenegger et al.}, 2000] would correspond to arcs
$10^{1}-10^{2}$ days long\footnote{However, as pointed out in
[\textit{Iorio}, 2001], short arcs would not allow to average out
many gravitational perturbations}. We could take advantage of this
fact by choosing the arcs so to avoid the eclipses effects;
moreover, over such short time spans it would be possible to
consider the satellite's spin axis direction as fixed in the
inertial space. Indeed, many thermal effects vanish for a suitable
fixed spin axis direction.

The basic assumptions of our study are the following: \\
$\bullet$ In order to simplify the calculations we will consider a
couple of \lg\ type satellites with \sa\ $a=12270$ km, as for \lg,
zero eccentricity $e$ and inclination $i$ and orbital
period $P=1.35\times 10^{4}$ s\\
$\bullet$ The physical parameters of the satellites like
area-to-mass ratio $S/m$, reflectivity and drag coefficients $C_R$
and $C_D$, etc. are assumed to be equal to those of \lg:
$S/m=6.87\times 10^{-3}$ \textrm{cm$^{2}$\ g$^{-1}$}, $C_R=1.13$,
$C_D=4.9$\\
$\bullet$ We will assume that their spins are aligned with the
Earth's spin axis assumed as $z$ axis of a terrestrial equatorial
inertial frame\\
$\bullet$ Only orbital arcs in full sunlight will be considered:
the eclipses effects will be neglected\\
$\bullet$ The perturbations on $r$ and $\varphi$ will be averaged
over an orbital revolution\\
$\bullet$ The perturbations which will be considered are the
direct solar radiation pressure, the Earth's albedo, the
Poynting-Robertson effect, the direct Earth IR radiation pressure,
the anisotropic thermal radiation or \yse, the thermal trust or
\re, the atmospheric drag and the effect of the Earth's magnetic
field.

The paper is organized as follows: in Section 2 and Section 3 the
perturbations on $r$ and $\varphi$, respectively, are worked out,
while Section 4 is devoted to the conclusions.

%--------------------------------
\section{The radial position}
The radial position of a satellite in a perturbed circular orbit
may change due to variations both in its \sa\ $a$ and its
eccentricity $e$ according to [{\it Christodoulidis et al.},
1988]: \eqi \Delta r=\sqrt{(\Delta a)^{2}+\frac{1}{2}(a\Delta
e)^{2}}.\lb{dr}\eqf The perturbations on $a$ are calculated with
[{\it Milani et al.,} 1987]: \eqi\dot a=\frac{2}{n}T,\lb{dadt}\eqf
where $n=\frac{2\pi}{P}$ is the mean motion, $P$ is the Keplerian
satellite's orbital period and $T$ is the along-track disturbing
acceleration. Concerning the eccentricity, since we are dealing
with circular orbits we will use the components of the
eccentricity vector $h=e\cos\omega$, $k=e\sin\omega$ ($\omega$ is
the argument of perigee) and [{\it Milani et al.,} 1987]: \eqi\dot
h \equiv\dot e\cos\omega+\mathcal{O}(e)=
\frac{1}{na}[-R\cos\ogf+2T\sin\ogf]+\mathcal{O}(e),\lb{ddh}\eqf
\eqi\dot k\equiv\dot e\cos\omega+\mathcal{O}(e)=
\frac{1}{na}[R\sin\ogf+2T\cos\ogf]+\mathcal{O}(e),\lb{ddk}\eqf
where $f$ is the true anomaly and $R$ is the radial component of
the perturbing acceleration. For a circular orbit, from \rfr{ddh}
and \rfr{ddk} it follows: \eqi \dot e =\sqrt{(\dot h)^{2}+(\dot
k)^{2}}.\eqf
%............................................
\subsection{The radiative perturbations}
The perturbing acceleration due to direct solar radiation pressure
can be approximately written as [{\it Lucchesi,} 1998]: \eqi{\bf
w}_{\odot}=-w_{\odot}\sv,\lb{arak}\eqf where $\sv$ is the unit
vector from the Earth to the Sun and:\eqi w_{\odot}
=\rp{S}{m}\rp{I_0}{c}C_R. \lb{solar}\eqf In \rfr{arak} the term
$(d_{\odot}/r_{\odot})^2$, in which $d_{\odot}$ is the \sa\ of the
Earth orbit around the Sun and $r_{\odot}$ is its instantaneous
distance, has been set equal to one. $I_0$ is the solar constant.
For \lg\ type satellites \rfr{solar} amounts to almost $3.6\times
10^{-7}$ cm s$^{-2}$.

Concerning the \sa, it turns out that if the total solar radiation
force acting on the spacecraft can be expressed in the general
form $\textbf{F}(\sv)$, no long-term effect in $a$ will appear to
any order in $e$ [\textit{Milani et al.,} 1987]. So, $\med{\dot
a}=0$.

Regarding the eccentricity, it can be proved that it is affected,
at zero order in $e$, by long-term periodic perturbations with an
almost yearly period [{\it Lucchesi,} 1998]. The amplitudes for
the components of the eccentricity vector are:
\eqi\med{\dot{h}}\propto \frac{3}{2}\frac{w_{\odot}}{na},\eqf
\eqi\med{\dot{k}}\propto \frac{3}{2}\frac{w_{\odot}}{na},\eqf so
that, over an orbital revolution, the radial position of the
satellite changes of almost 11 cm [\textit{Milani et al.,} 1987].

Assuming that the solar constant is known at $0.3\%$
[\textit{Ciufolini et al.}, 1997], the major source of uncertainty
in the satellite's radial position due to the direct solar
radiation pressure would reside in the $C_R$ satellite's
coefficient. Assuming a global $0.5\%$ mismodeling, the systematic
error would amount to $\delta r_{SRP}=5\times 10^{-2}$ cm. This
result suggests that, in order to meet the requirement of $\delta
r\sim 10^{-2}$ cm, a careful analysis of the optical properties of
the satellites should be conducted in accurate pre-launch
controls.

For a circular and equatorial orbit the perturbations on the
satellite's radial position induced by the Earth's albedo are
entirely due to the changes in the eccentricity vector as well.
Indeed, the along-track acceleration, which is particularly
sensitive to the specular part of Earth's albedo and to its
spatial and temporal variations, can be expanded in terms of
long-periodical harmonics whose coefficients are proportional to
$\sin i$ [\textit{Anselmo et al.}, 1983]. Concerning the
eccentricity vector, the simple analytic model by
\textit{M\'{e}tris et al.,} [1997], based on an uniform mean
albedo of $\overline{\mathcal{A}}=0.3$, turns out to be adequate.
The related perturbation amounts to $8\times 10^{-2}$ cm per
orbit. Then, the systematic error due to the mismodeling in
$\overline{\mathcal{A}}$, assumed to be $10\%$ [\textit{Lucchesi},
1998], amounts to  $8\times 10^{-3}$ cm per orbit.

It may be important to note that a way to reduce the impact of the
direct solar radiation pressure and the albedo could be the use of
a couple of dense satellites with small $\frac{S}{m}$.

The Poynting-Robertson acceleration [\textit{Burns et al.,} 1979]
leaves unaffected the eccentricity while changes the \sa\ with
long-term perturbations whose nominal amplitudes are of the order
of $10^{-3}-10^{-4}$ cm per orbit. Consequently, the related
systematical errors are negligible.
%..................................
\subsection{The thermal perturbations}
Concerning the direct Earth IR radiation pressure, the Earth
thermal emissivity ${\mtc{E}}$ can be modelled in a form of a
latitude-dependent spherical harmonic expansion on a spherical
Earth surface [{\it Sehnal}, 1981]: \eqi
{\mtc{E}}={\mtc{E}}_0+{\mtc{E}}_1(t)\ P_1(\sin{\f})+{\mtc{E}}_2\
P_2(\sin{\f})+...,\eqf where $\f$ is the terrestrial latitude.
Regarding the effects of the first two zonal constant terms
${\mtc{E}}_0$ and ${\mtc{E}}_2$, $\med{\dot a}=0$ since it is
proportional to $(e\sin i)^{2}$ and $\med{\dot e}=0$ since it is
proportional to $e(\sin i)^{2}$ [{\it Sehnal}, 1981]. The
perturbations due to ${\mtc{E}}_1(t)$, which shows an
approximately yearly variation, vanish as well since $R$ and $T$
are proportional to $\sin i$ [\textit{M\'{e}tris et al.}, 1997].

The \yse\ [\textit{Afonso et al.,} 1989] induces perturbations on
the semimajor axis which
 vanishes since the first non-zero term in $\dot a$
is of order $\mathcal{O}(e)$ for orbital arcs in full sunlight. As
far as the eccentricity vector is concerned, $e$ would be
perturbed at a level of $9\times 10^{-2}$ cm per orbit. However,
such effect depends on the orientation of the satellite's spin
axis $\hat \xi$: for a fixed direction such that $\xi_x=\xi_y=0$,
as it could be feasible over short arcs, $\med{\dot h}=\med{\dot
k}=0$ [\textit{Lucchesi}, 1998].

The thermal thrust or \re\  [\textit{Rubincam,} 1987] depends on
$\hat\xi$ as well; for $\xi_x=\xi_y=0$ it vanishes because it can
be proved that, in this case, $R$ and $T$ are proportional to
$\sin i$.
%..........................................
\subsection{Other perturbations}
The neutral and charged drag has negligible effects on $r$.
Indeed, by neglecting the corotation of the exosphere, it turns
out that [\textit{Milani et al.,} 1987]:\eqi \med{\Delta a}=-C_D
\frac{S}{m}a^{2}\varrho,\eqf in which $\varrho$ is the atmospheric
density. For $\varrho=8.4\times 10^{-21}$ g cm$^{-3}$
[\textit{Afonso et al.}, 1985] we would have a decay in $a$ of
$4.2\times 10^{-4}$ cm per revolution. The same holds for the
charged drag [\textit{Rubincam}, 1990]. For circular orbits the
eccentricity is not affected by the drag [\textit{Milani et al.,}
1987].

If the satellites carry an electric charge, as it is the case for
\lg, the Earth's magnetic field acts upon them via the Lorentz
force. Its effect on the \sa\ is of course zero since the Lorentz
force does not change the satellite's total mechanical energy $W$
and, consequently, $a$: indeed, $W=-\frac{GM}{2a}$. Regarding the
eccentricity vector, it turns out that $\med{\dot h}=\med{\dot k
}=0$ because $R$ is constant and $T$ is of order $\mathcal{O}(e)$.
%--------------------------------
\section{The azimuthal position}
As in [\textit{Iorio,} 2001; \textit{Lichtenegger et al.,} 2001],
for a satellite in an equatorial and circular orbit the rate of
the azimuthal angle  can be calculated by means of: \eqi
\dot\varphi=\dot\omega+\dot\Omega\cos i+\dot
\mathcal{M}.\lb{dfdt}\eqf In \rfr{dfdt} $\Omega$ is the longitude
of the ascending node and $\mathcal{M}$ is the mean anomaly whose
rate equation is given by: \eqi \dot
\mathcal{M}=n-\frac{2}{na}R\frac{r}{a}-\sqrt{1-e^{2}}(\dot\omega+\cos
i\dot\Omega).\lb{dmdt}\eqf For $e=i=0$ \rfrs{dfdt}{dmdt} yield:
\eqi\dot\varphi=n-\frac{2}{na}R.\lb{dfdt2}\eqf In dealing with
\rfr{dfdt2} the indirect effects on $n$ induced by the
perturbations on $a$ must be considered as well. At the first
perturbative order they are: [\textit{Milani et al.,} 1987] \eqi
\Delta n=-\frac{3n}{2a}\Delta a.\eqf
%...............................
\subsection{The radiative perturbations}
The direct solar radiation pressure does not affect $\varphi$
because, as seen in previous section, $\med{\dot a}=0$; moreover,
$\med{R}=0$.

Regarding the Earth's albedo, the indirect perturbations on $n$
vanish because $\med{\dot a}=0$. According to the model by
\textit{M\'{e}tris et al.,} [1997] $\med{R}=0$.

The non-vanishing perturbations of order $\mathcal{O}(e^{0})$
induced by the Poynting-Robertson effect, which amount to almost
$10^{-5}$ mas per revolution or less are negligible.
%..................................
\subsection{The thermal perturbations}
Concerning the direct Earth IR radiation pressure, $\med{\dot
a}=0$ so that the indirect perturbations on the mean motion vanish
as well. The first two constant zonal terms of the Earth IR
emissivity yield non-vanishing terms of zero order in $e$. Indeed,
it turns out that, for $\mathcal{E}_{0}$: \eqi
\med{-\frac{2}{na}R}^{(0)}=-\frac{2}{na}(\frac{R_{\oplus}}{a})^{2}\mathcal{E}_{0}
\frac{S}{m}\frac{C_R}{c},\eqf which yields $-1.6\times 10^{-1}$
mas per revolution. The systematic error induced by the
mismodeling in $\mathcal{E}_{0}$ is within the limit
$\delta\varphi\sim 10^{-2}$ mas per revolution and could be
reduced by using satellites with small area-to-mass ratio. The
bias  due to $\mathcal{E}_{2}$ is negligible since its nominal
perturbation amounts to $1\times 10^{-3}$ mas per revolution. As
in for $r$, also in this case $\mathcal{E}_{1}(t)$ does not
contribute since $R$ is proportional to $\sin i$.

The \yse\ does not affect the azimuthal position since $\med{\dot
a}=0$ and it turns out that $R$ averages out over an orbital
revolution. The same holds for the \re\ which is not present when
$\xi_z=\pm 1$ and $i=0$.
%..........................................
\subsection{Other perturbations}
The indirect perturbation on $n$ due to the drag shift experienced
by the  \sa\ is negligible because it amounts to $6\times 10^{-4}$
mas per revolution, while: \eqi
\med{-\frac{2}{na}R}=C_D\frac{S}{m}nae\med{\sin E}=0,\eqf where
$E$ is the eccentric anomaly.

The effect of the Earth's magnetic field is completely negligible
since it is of the order of $10^{-5}$ mas per revolution.

The major source of error in the azimuthal position turns out to
be the Earth's $GM$. Indeed, $\Delta\varphi ^{(GM)}=\frac{\delta
(GM)}{2na^{3}}\times P=1.2$ mas per revolution by assuming $\delta
(GM_{\oplus})=8\times 10^{11}$ cm$^{3}$ s$^{-2}$
[\textit{McCarthy}, 1996].

%--------------------------------
\section{Conclusions}
In the context of the gravitomagnetic clock effect, by considering
a couple of identical SLR satellites of \lg\ type with $e=i=0$ and
$\xi_{x}=\xi_{y}=0$, $\xi_{z}=\pm 1$ over short orbital arcs in
full sunlight, it turns out that some non-gravitational
perturbations on $r$ and $\varphi$ vanish.

Regarding the radial position, the largest perturbations are due
to the direct solar radiation pressure which would change the
satellite's distance of almost $10^{1}$ cm per revolution and the
Earth's albedo which would induce a change of $8\times 10^{-2}$ cm
per revolution. However, their systematic errors induced by the
mismodeling in the optical properties of the satellites and the
Earth's albedo should fall below the cutoff $\delta r\sim 10^{-2}$
cm. The influence of the other non vanishing perturbations, like
Poynting-Robertson effect and neutral and charged drag, is
negligible. The thermal perturbations vanish.

The azimuthal angle is perturbed by the Earth's IR direct
radiation pressure at a level of $10^{-1}$ mas per revolution. The
other non vanishing perturbations are negligible.

It should be pointed out that all the non-vanishing perturbations
are proportional to $\frac{S}{m}$. This means that the impact of
their systematic errors could be reduced by using particularly
dense satellites with small area-to-mass ratio. However, the
largest source of error in $\varphi$ is the uncertainty in the
Earth's $GM$ which induce a bias of 1.2 mas per revolution. This
is a hard limitation to overcome because it is independent of the
particular satellite employed and is related to our knowledge of
the terrestrial gravitational field. Moreover, as pointed out in
[\textit{Iorio,} 2001], at the present level of knowledge of the
terrestrial gravitational field, the mismodeled gravitational time
varying perturbations induce systematic errors which are larger
than the required constraints; they could be overcome by averaging
them over adequately long time spans.

The results of the present work suggest that for a suitable choice
of the satellites to be employed and of their orbital geometry it
would be possible to keep the non-gravitational perturbations
within the required constraints in order to make feasible the
measurement of the gravitomagnetic clock effect. As far as the
systematic errors induced by the forces acting upon the satellites
are concerned, the major problems come from the Earth's
gravitational environment. Improvements in satellite tracking
accuracy of almost two orders of magnitude are needed as well.
%--------------------------------
\section*{Acknowledgements}
I thanks L. Guerriero for his support and encouragement.

%%%%%%%%%%%%%%%%%%%%%%%%%%%%%%%%%

\end{document}